\documentclass[printer]{raa}
\usepackage{natbib}
\usepackage{graphicx,times}
\graphicspath{{figs}}
\usepackage{amssymb,amsmath,mathrsfs}
\usepackage{multirow}
\usepackage{booktabs,threeparttable}
\bibpunct{(}{)}{;}{a}{}{,}
\usepackage[pagebackref=true]{hyperref}
\usepackage{rotating}
\usepackage{enumitem}
\usepackage{orcidlink}

\newcommand{\bea}{\begin{eqnarray}}
\newcommand{\eea}{\end{eqnarray}}

\usepackage{ulem}

\begin{document}

\title{Optical Continuum Light Curves and Bolometric Energy Estimates of Solar White-light Flares}

\volnopage{ {\bf 2026} Vol., {\bf X} No. {\bf XX}, 000--000}
\setcounter{page}{1}

\author{Yingjie Cai\inst{1,2}\orcidlink{0009-0007-4469-0663}, Yijun Hou\inst{1,2}\thanks{E-mail: yijunhou@nao.cas.cn}\orcidlink{0000-0002-9534-1638}, Hengkai Ding\inst{4,5}\orcidlink{0000-0003-3787-1929}, Ting Li\inst{1,2}\orcidlink{0000-0001-6655-1743}, Jifeng Liu\inst{2,3,6,7}\orcidlink{0000-0002-2874-2706}}
  \institute{
   $^1$ State Key Laboratory of Solar Activity and Space Weather, National Astronomical Observatories, Chinese Academy of Science, Beijing 100101, China\\
   $^2$ School of Astronomy and Space Science, University of Chinese Academy of Sciences, Beijing 100049, China\\
   $^3$ Key Laboratory of Optical Astronomy, National Astronomical Observatories, Chinese Academy of Sciences, Beijing 100101, China\\
   $^4$ Purple Mountain Observatory, Chinese Academy of Sciences, Nanjing 210023, China\\
   $^5$ School of Astronomy and Space Sciences, University of Science and Technology of China, Hefei 230026, China\\
   $^6$ Institute for Frontiers in Astronomy and Astrophysics, Beijing Normal University, Beijing 100875, China\\
   $^7$ New Cornerstone Science Laboratory, National Astronomical Observatories, Chinese Academy of Sciences, Beijing 100101, China\\
  }

\vs \no
{\small Received  ; accepted  }

\abstract{
Solar white-light flares (WLFs) are solar flares exhibiting enhanced emission in the optical continuum. They are critical for understanding energy release and transport mechanisms in solar flares and for conducting comparative studies with stellar WLFs. However, the scarcity of accurately and reliably measured optical continuum light curves for solar WLFs significantly hampers related studies. Based on the optimized solar WLF identification method, we construct a dataset of optical continuum light curves for 70 solar WLFs using 6173 {\AA} continuum intensity images from the Solar Dynamics Observatory. Moreover, for each solar WLF event, we also provide the location of the white-light emission enhancement signals and key parameters including bolometric energies and durations derived from both the traditional fixed-temperature blackbody model and the refined variable-temperature blackbody model. This dataset will serve as a valuable resource for future statistical investigations of solar WLFs and for comparative studies between solar and stellar flares.
}

 \authorrunning{Yingjie Cai et al.}    
 \maketitle

\section{Introduction}
\label{sec:introduction}
Solar flares are intense bursts of electromagnetic radiation originating from the rapid release of magnetic energy in the localized solar atmosphere. In some cases, solar flares also exhibit an apparent emission enhancement in the optical continuum, in addition to that typically observed in extreme ultraviolet (EUV) and X-ray channels. These events are known as solar white-light flares (WLFs) \citep{1970SoPh...13..471S, 1989SoPh..121..261N, 1993SoPh..144..169N}. Distinguished from enhanced emission at EUV and X-ray wavelengths, the strong white-light (WL) emission enhancement signifies that a substantial amount of energy is deposited in the lower solar atmosphere (photosphere and lower chromosphere), which makes solar WLFs critical for understanding the mechanisms of energy transport and heating during solar flares \citep{1995SoPh..157..271F, 2003A&A...403.1151D, 2017NatCo...8.2202H, 2018A&A...613A..69S, 2018ApJ...867..159S, 2020ApJ...893L..13S, 2023ApJ...954....7L, 2025ApJ...979L..43Y, 2025ApJ...986L..15X}. Beyond the Sun, WLFs are also routinely observed on other stars, whose energies are 10$^{1}$-10$^{4}$ times larger than those of the maximum solar flares \citep{2012Natur.485..478M, 2017ApJ...851...91N}. As stellar flare observations lack spatial resolution, solar WLFs serve as the Rosetta Stone for interpreting the physics of much more energetic stellar WLFs. Detailed studies of solar WLFs are therefore indispensable for bridging the gap between spatially resolved solar flares and spatially unresolved stellar flares.

\begin{figure*}[htbp]
\centering
\includegraphics [width=1\textwidth]{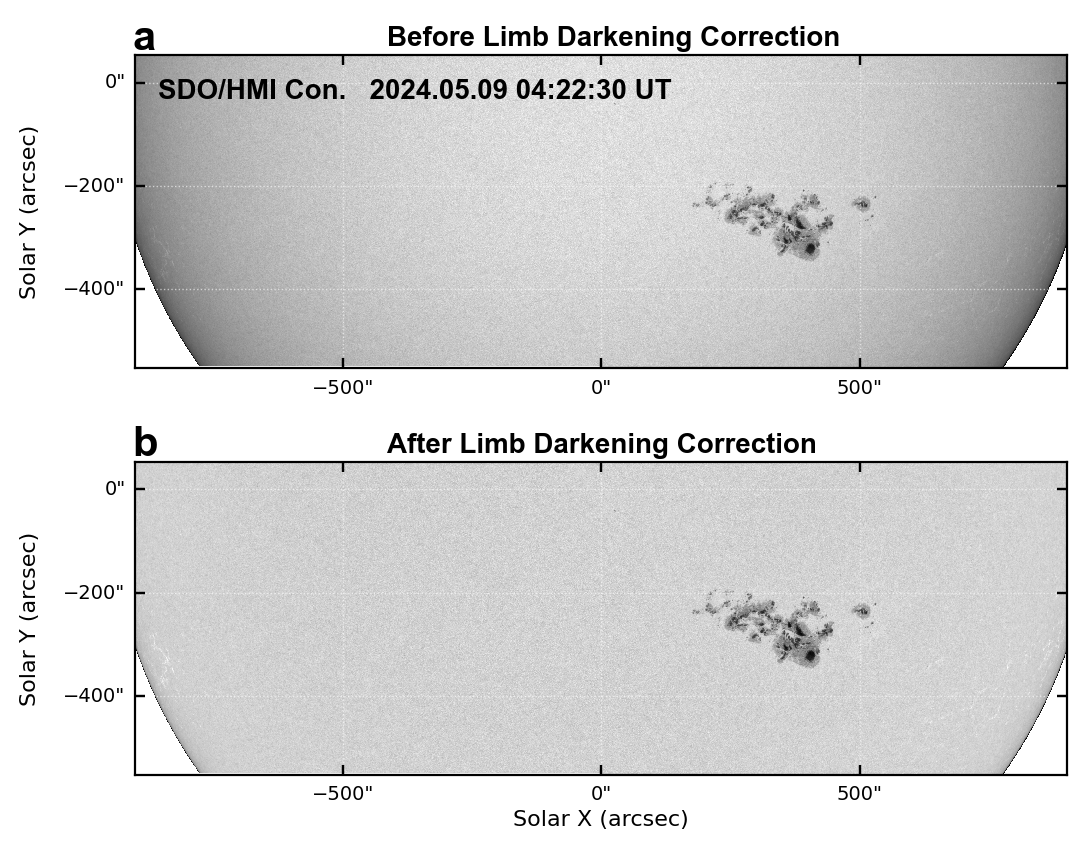}
\caption{Demonstration of limb-darkening correction for SDO/HMI 6173 {\AA} continuum intensity images. (a) Original solar continuum intensity image observed by SDO/HMI at 04:22:30 UT on 09 May 2024. (b) The corresponding image after the limb-darkening correction has been applied.}
\label{fig1}
\end{figure*}

Despite significant advances in our understanding of solar WLFs \citep{2018ApJ...867..159S, 2023ApJ...952L...6S, 2024ApJ...963L...3L, 2024SoPh..299...11J, 2024ApJ...975...69C, 2025ApJ...992...72J, 2025ApJ...979L..43Y, 2025ApJ...986L..15X} based on observations from Goode Solar Telescope (GST) \citep{2010AN....331..636C}, the New Vacuum Solar Telescope (NVST) \citep{2014RAA....14..705L}, the Solar Dynamics Observatory (SDO) \citep{2012SoPh..275....3P}, the Chinese H{$\alpha$} Solar Explorer (CHASE) \citep{2022SCPMA..6589602L} and the Advanced Space-based Solar Observatory (ASO-S) \citep{2023SoPh..298...68G}, a scarcity of accurately measured optical continuum light curves for solar WLFs remains. This scarcity not only hampers statistical studies of key solar WLF parameters but also critically limits robust comparisons with the abundant optical continuum light curves of stellar WLFs. To address this critical gap, we construct and present a dataset of optical continuum light curves for 70 solar WLFs, based on observations from the Helioseismic and Magnetic Imager (HMI) \citep{2012SoPh..275..207S} and the Atmospheric Imaging Assembly (AIA) \citep{2012SoPh..275...17L} aboard SDO. This dataset provides a foundational resource designed to enable future statistical inquiries into solar WLF energetics and to facilitate direct comparisons with stellar WLFs.

The structure of this paper is organized as follows. Section \ref{sec:observations} presents the observations of 70 solar WLFs from SDO, including HMI 6173 {\AA} continuum intensity images and AIA 1600 {\AA} images. Section \ref{sec:methods} outlines the data processing methodology, encompassing data preprocessing, identification of WL emission enhancement signals, acquisition and detrending of optical continuum light curves and calculation of bolometric energy and duration. Section \ref{sec:released data} presents a description of the data released in this study. Ultimately, in Section \ref{sec:future}, we discuss the potential role of this dataset in future studies of both solar and stellar WLFs.

\begin{figure*}[htbp]
\centering
\includegraphics [width=1\textwidth]{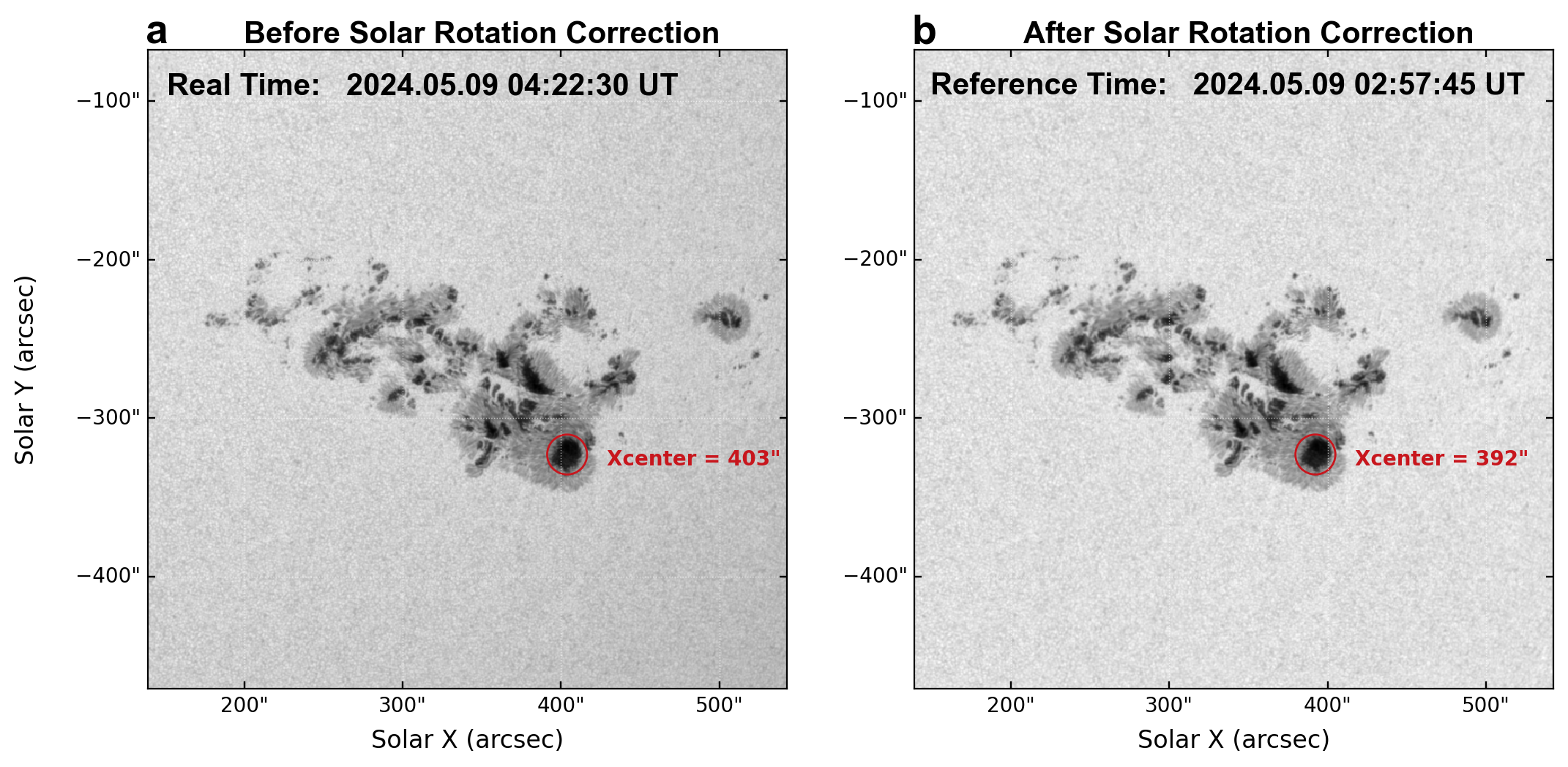}
\caption{Demonstration of solar rotation correction on SDO/HMI 6173 {\AA} continuum intensity images. (a) Original solar continuum intensity image observed by SDO/HMI at 04:22:30 UT on 09 May 2024. (b) The corresponding image after being differentially rotated to a common reference time of 02:57:45 UT on the same day. The red circles in both panels highlight the same feature region.}
\label{fig2}
\end{figure*}

\section{Observations}
\label{sec:observations}

The primary data used in this study are from HMI and AIA on board SDO. SDO/HMI observes the Sun using the Fe I 6173.3 {\AA} spectral line and records intensities at six wavelength points across this absorption line, from which the continuum intensity (Ic) is derived. We utilize the level-1.5 data product hmi.Ic$\_45s$, which provides full-disk continuum intensity maps with a cadence of 45 s and a spatial resolution of $0.5^{\prime \prime}$ pixel\textsuperscript{-1}. The HMI continuum data provides a map of the photospheric emission, which is essential for detecting WL emission enhancement during flares. SDO/AIA is designed to capture comprehensive images of the solar atmosphere, which images the full Sun in 10 different wavelength channels, including EUV, ultraviolet (UV) and visible passbands. For this study, we specifically use the 1600 {\AA} images, which have a cadence of 24 s and a spatial pixel resolution of $0.6^{\prime \prime}$. The AIA 1600 {\AA} data is primarily used to accurately identify the flare ribbon regions, which serve as the primary areas for searching for WL emission enhancement signals.

The sample of solar WLFs analyzed in this study is drawn from two sources: (1) a long-term sample covering February 2011 to July 2023 \citep{2024ApJ...975...69C}, which contributes 39 WLFs (GOES class $\geq$ C5.0) located within 75$^\circ$ of the central meridian; and (2) observations of the super active region (SAR) NOAA 13664/13697 between 2 May and 9 June 2024 (Cai et al. 2026, in preparation), which provide 31 WLFs (GOES class $\geq$ M1.0) located within 90$^\circ$ of the disk center. From the initial candidates provided by these two sources, the final 70 events (33 X-class, 32 M-class, and 5 C-class) were selected based on the quality and completeness of their optical continuum light curves. To ensure accurate derivation of physical parameters like bolometric energy and duration, we specifically excluded cases with data gaps or complex background variations.

\begin{figure*}[htbp]
\centering
\includegraphics [width=0.85\textwidth]{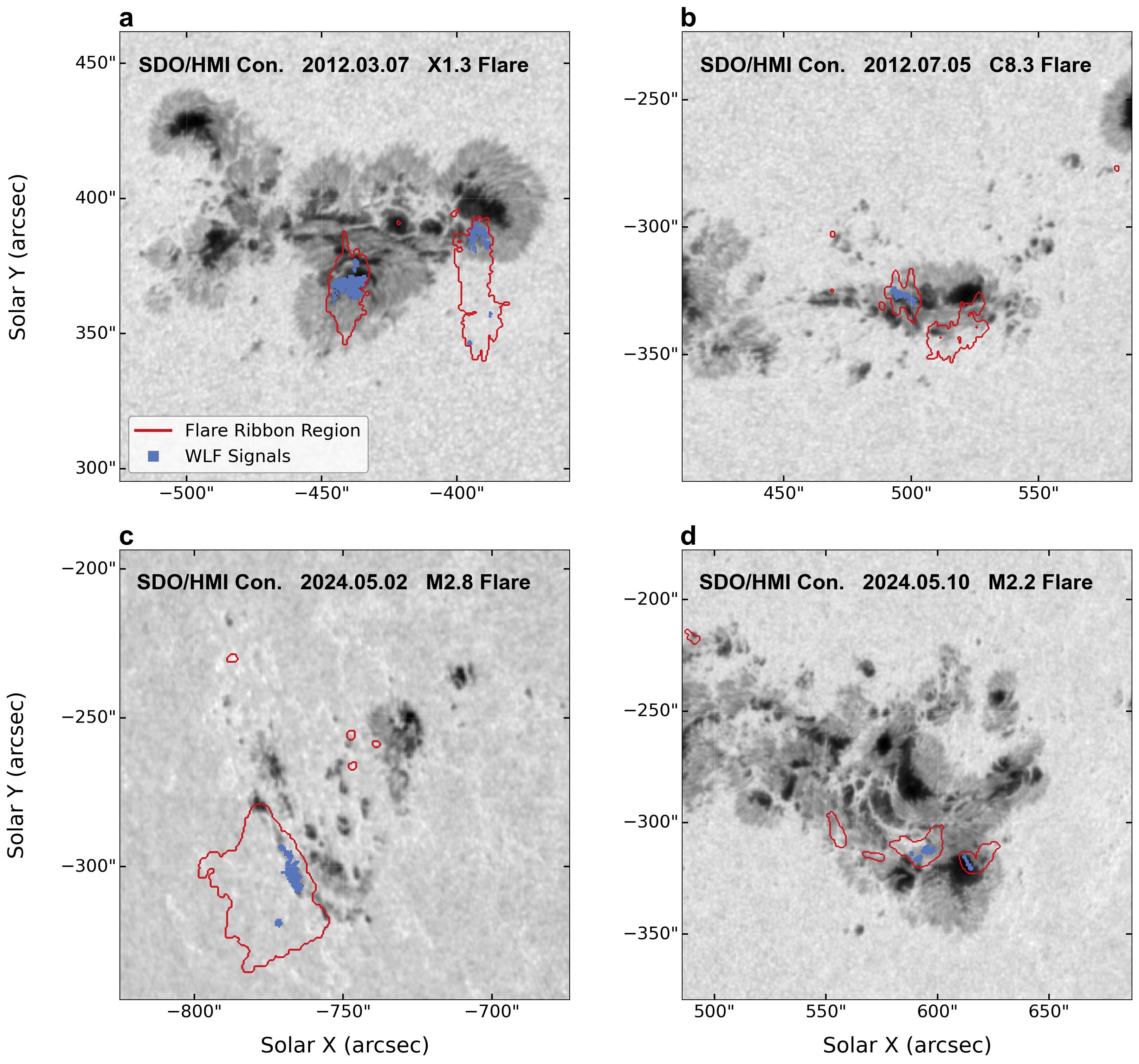}
\caption{Spatial distributions of WL emission enhancement signals within flare ribbon regions for four representative solar WLFs. (a)-(d): SDO/HMI 6173 {\AA} continuum intensity images for an X1.3-class solar WLFs on 2012 March 7, a C8.3-class solar WLFs on 2012 July 5, an M2.8-class solar WLFs on 2024 May 2, and an M2.2-class solar WLFs on 2024 May 10, respectively. In each panel, the flare ribbon regions, as identified from SDO/AIA 1600 {\AA} observations, are outlined in red. The detected WL emission enhancement signals (WLF signals) within these flare ribbon regions are highlighted by the blue patches.}
\label{fig3}
\end{figure*}

\section{Methods}
\label{sec:methods}

    \subsection{Data Preprocessing}
    \label{subsec:3.1}

    The raw observational data of HMI and AIA were obtained from the website (\url{http://jsoc.stanford.edu/ajax/lookdata.html}) of Stanford SDO Joint Science Operations Center (JSOC). Firstly, we determined the locations of these solar flares and cropped the original full-disk images of the Sun accordingly through a visual inspection of the high-cadence movies using the JHelioviewer software. Subsequently, the cropped data of HMI and AIA underwent preprocessing steps, respectively.

\begin{figure*}[htbp]
\centering
\includegraphics [width=0.72\textwidth]{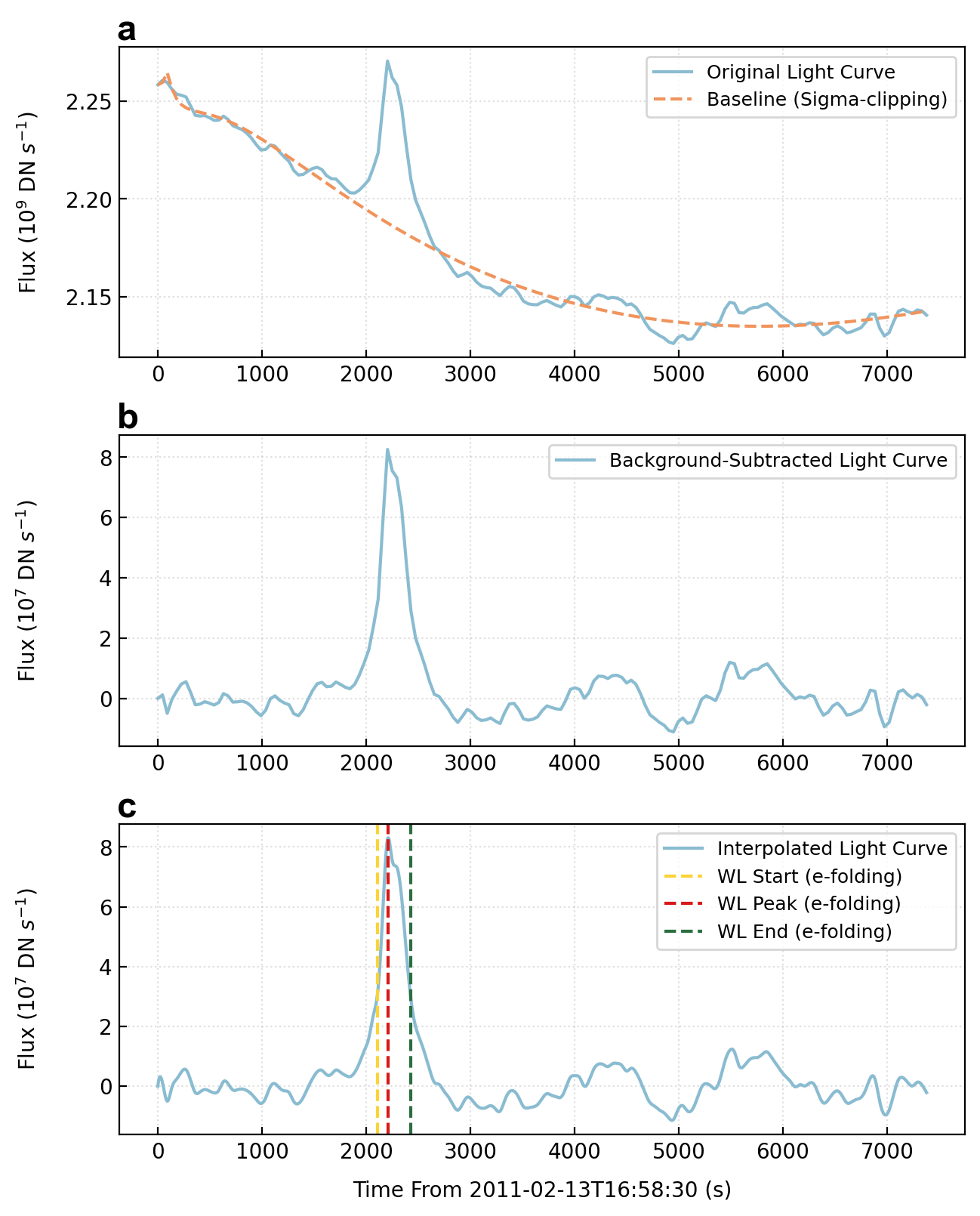}
\caption{Data processing pipeline for extracting solar WLF parameters from a SDO/HMI 6173 {\AA} continuum light curve. (a) The original optical continuum light curve (blue solid line) with its baseline (orange dashed line) derived by a iterative sigma-clipping method. (b) The background-subtracted optical continuum light curve obtained by removing the baseline from the original data. (c) The interpolated optical continuum light curve used for precise measurement of bolometric energy and duration for each solar WLF. The vertical dashed lines mark the start, peak, and end times of the solar WLF, determined using the e-folding method.}
\label{fig4}
\end{figure*}

    To ensure the accuracy and reliability of the subsequent analysis, all HMI 6173 {\AA} continuum intensity images underwent two essential preprocessing steps: (1) limb-darkening correction; and (2) solar rotation correction. The observed intensity across the solar disk is non-uniform, exhibiting a characteristic dimming towards the limb (limb darkening) in the visible continuum observed by HMI at 6173 {\AA}. This center-to-limb intensity variation, if left uncorrected, would introduce significant systematic errors in the absolute photometric measurements of WL emission, consequently affecting the accurate determination of key parameters such as bolometric energy. Therefore, we applied a second-order limb-darkening correction to all HMI 6173 {\AA} continuum intensity images \citep{2020ApJ...904...96C}. This correction provides a uniform background essential for precise measurement of the WL emission. The corrected 6173 {\AA} continuum intensity of each pixel is given by
    \begin{equation}
    \ I_{ij}^{corr} = \frac{I_{ij}^{non - corr}}{C_{ij}}, \label{eq:1}
    \end{equation}
    where $C_{ij}$ refers to the limb darkening function
    \begin{equation}
    \ C_{ij} = 1-u_\lambda-\nu_\lambda+u_\lambda\cos(\Theta)+\nu_\lambda\cos(\Theta)^2,  \label{eq:2}
    \end{equation}
    where $\Theta = \sin^{-1}(\sqrt{(x_{i}-x_{c})^{2}+(y_{j}-y_{c})^{2}}/R_{\odot})$. $(x_{i},y_{i})$ and $(x_{c},y_{c})$ respectively refer to the given pixel coordinates and coordinates of the solar disk center. Moreover, the coefficients $u_\lambda$ and $\nu_\lambda$ in Equation \ref{eq:2} are wavelength-dependent. For the HMI 6173 {\AA} continuum intensity images, the values of these coefficients are respectively equal to $u_{6173.3}=0.836$ and $\nu_{6173.3}=-0.204$ \citep{1976asqu.book.....A}. As shown in Figure \ref{fig1}, we display the demonstration of limb-darkening correction for HMI 6173 {\AA} continuum intensity images. Obviously, the correction process successfully removes the center-to-limb intensity variation, resulting in a more uniform brightness across the entire disk of the Sun.

    Furthermore, the apparent drift of solar features from east to west due to solar differential rotation also poses a challenge for tracking the intrinsic evolution of a target region. To mitigate this effect, we aligned the time series of HMI 6173 {\AA} continuum images during a solar WLF event to a common reference time. This co-alignment process effectively removes the apparent motion caused by solar rotation, ensuring that the temporal evolution of the continuum intensity at a fixed location reflects genuine physical changes within the flare ribbon region rather than geometric effects. Figure \ref{fig2} illustrates the solar rotation correction on HMI 6173 {\AA} continuum intensity images. Figure \ref{fig2}(a) and \ref{fig2}(b) display the same solar image observed at 04:22:30 UT on 09 May 2024. The key distinction is that Figure \ref{fig2}(a) presents the data at its original observation time, while Figure \ref{fig2}(b) shows the result after the image has been differentially rotated to a common reference time of 02:57:45 UT on the same day. It can be seen that the feature (marked by the red circle) centered at an X-coordinate of approximately $403^{\prime \prime}$ at the original time of 04:22:30 UT on 09 May 2024, has shifted to $392^{\prime \prime}$ at the reference time of 02:57:45 UT.

    Similarly, the AIA 1600 {\AA} data undergo the same solar differential rotation correction procedure to achieve precise co-alignment with the processed HMI 6173 {\AA} continuum intensity images in both space and time. These co-aligned AIA 1600 {\AA} images were then used to identify the flare ribbon regions. This was achieved by taking the union of the brightened areas observed in each image during flares in the AIA 1600 {\AA} passband. The resulting flare ribbon regions served as the primary search areas for the subsequent identification of WL emission enhancement signals.

    \subsection{Identification of White-light Emission Enhancement Signals}
    \label{subsec:3.2}

    The identification of WL emission enhancement signals was performed on the preprocessed HMI 6173 {\AA} continuum intensity images, following the methodology established by \citet{2024ApJ...975...69C}. A pixel is identified as a valid WL emission enhancement signal only when at least nine of its eighteen associated pixels (the central pixel and its surrounding pixels at two consecutive moments) exhibit WL emission enhancement ($\delta_n$) larger than the intrinsic threshold. The WL emission enhancement ($\delta_n$) for each pixel at every moment is calculated by
    \begin{equation}
    \delta_{n} = \left|\frac{I_{n+1}-I_{n}}{I_{n}}\right|, \label{eq:3}
    \end{equation}
    where $I_{n}$ and $I_{n+1}$ represent the WL continuum intensity at the adjacent moments. The intrinsic threshold for each pixel is defined as the maximum value of its $\delta_n$ during the 30 minutes preceding the flare onset. It is worth noting that a fundamental premise of our identification method is the established spatial correlation between WLFs and the flare ribbon regions observed by AIA 1600 {\AA}. Therefore, to enhance the reliability of detection and minimize false positives, the identification algorithm was applied exclusively within the spatial boundaries of the AIA 1600 {\AA} flare ribbon regions. Additionally, to further refine the selection of the identified WL emission enhancement signals, we applied an additional constraint based on the concept of connected domains. Only candidate pixels that formed connected domains of at least 9 pixels at a single time step were retained. The application of the above identification methodology is demonstrated in Figure \ref{fig3}, which presents the spatial distribution of the detected WL emission enhancement signals for four representative solar WLFs spanning different GOES classes.

    \subsection{Detrending of Optical Continuum Light Curves}
    \label{subsec:3.3}

    Following the identification of WL emission enhancement signals as described in Section \ref{subsec:3.2}, the original optical continuum light curves were extracted for subsequent analysis. For every solar WLF, the total continuum intensity at each moment was computed by summing the flux from all WL emission enhancement signals within its boundary at each time step. This yielded the original optical continuum light curve, which contains both the rapid flare-associated WL emission enhancement and the slower-varying background trend.

    To ensure statistical reliability and minimize bias arising from different detrending techniques, we aimed to apply a consistent background subtraction method across the entire dataset. For the vast majority of our sample (over 60 of the 70 events), we successfully employed the iterative sigma-clipping method \citep{2014ApJ...797..122D} to determine the background baseline using extended pre- and post-flare data. This method is particularly effective when continuous observations are available for a sufficient duration before and after the flare and no other flares occur during this period. The algorithm works by iteratively calculating the mean and standard deviation of the optical continuum light curve, then identifying and excluding data points that deviate significantly (i.e., outliers corresponding to the flare impulse) from the local mean. This process dynamically fits a background trend that is robust against the flare signal itself enabling accurate isolation of the net flare emission enhancement. The detrended optical continuum light curve ($L_{WLF}$) is obtained by subtracting this fitted background trend ($L_{baseline}$) from the original data ($L_{origin}$). Figure \ref{fig4} illustrates the procedure of the detrending process based on the iterative sigma-clipping algorithm.

    For the few remaining complex events where the pre-flare or post-flare backgrounds of flares are contaminated by other eruptive activity (e.g., in super active regions with frequent flaring), the sigma-clipping method becomes unreliable due to the inability to define a stable quiescent background. In such cases, we applied an empirical method based on local interpolation \citep{2017ApJ...851...91N, 2024ApJ...975...69C}. To objectively determine the start and end times for this interpolation, we prioritized the use of a standard $3\sigma$ threshold. Specifically, we calculated the mean ($\mu$) and standard deviation ($\sigma$) of the pre-flare background and defined the interpolation anchor points ($t_1$ and $t_2$) as the first and last moments, respectively, when the flux exceeded the $\mu + 3\sigma$ level. This approach proved effective for a subset of these events. However, we also encountered instances where the $3\sigma$ method remained inapplicable due to the intrinsic evolving properties of the flaring active regions. In these cases, the pre-flare level was already significantly elevated ($\mu$) or exhibited violent fluctuations ($\sigma$) due to prior heating or magnetic activity. As a result, the computed threshold ($3\sigma+\mu$) becomes unphysically high, driven by both the intense background and large variance. For these events, we resorted to identifying the local flux minima to define the flare interval. The standard approach is to select the first local flux minimum ($t_1$) before the flare peak and the first one ($t_2$) after the flare peak. This interval [$t_1$, $t_2$] is typically valid for simple, single-peaked optical continuum light curves. However, for complex, multi-peaked optical continuum light curves, visual inspection is required to select more appropriate minima. This may involve choosing a minimum earlier than $t_1$ or later than $t_2$, ensuring that the selected interval accurately represents the total flare period. Once the time interval [$t_1$, $t_2$] was determined (either via the $3\sigma$ threshold or local minima), a linear interpolation was performed between the flux values at these two points to model the underlying background trend. This interpolated segment replaced the original data within the flare interval. A smoothing filter (e.g., a boxcar filter over five data points) was applied to the entire reconstructed optical continuum light curve to generate a continuous baseline ($L_{baseline}$). The final detrended optical continuum light curve (($L_{WLF}$)), was calculated as $L_{WLF} = L_{origin} - L_{baseline}$.

    The choice between these distinct detrending methods was determined on a case-by-case basis by visually inspecting the optical continuum light curve of each flare, ensuring that the background trend was estimated as accurately as possible. The resulting detrended optical continuum light curves were used for subsequent calculations of bolometric energies and durations for solar WLFs.

\begin{figure*}[htbp]
\centering
\includegraphics [width=0.7\textwidth]{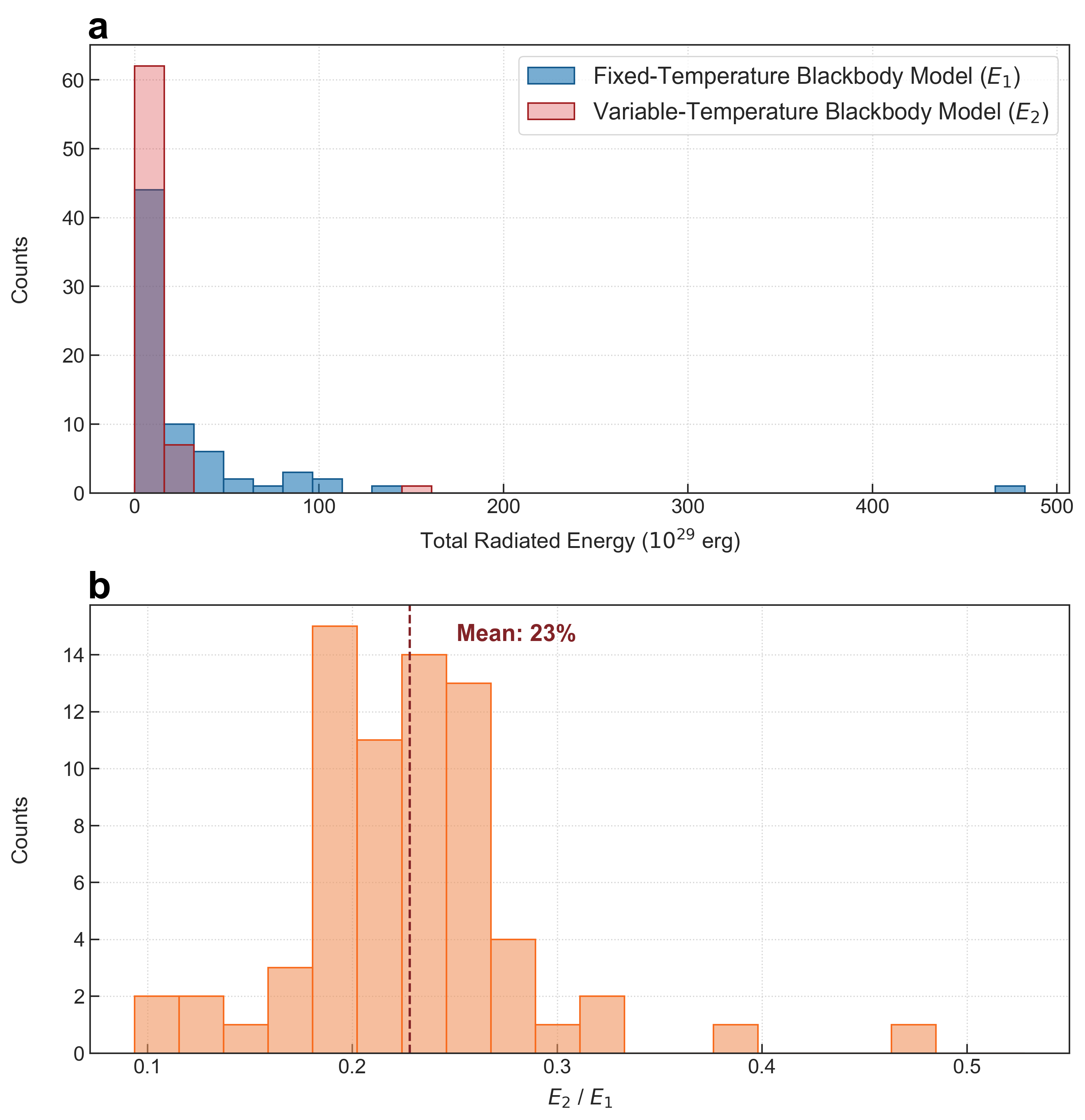}
\caption{Comparison of bolometric energies derived from the two calculation methods. (a) Distributions of bolometric energies for solar WLFs calculated using the fixed-temperature blackbody model and the refined variable-temperature blackbody model. (b) Distribution of the ratio between the two bolometric energies ($E_2 / E_1$).}
\label{fig5}
\end{figure*}

    \subsection{Calculation of Bolometric Energy and Duration}
    \label{subsec:3.4}

        \subsubsection{Bolometric Energy Calculation based on Fixed-Temperature Blackbody Model}
        \label{subsec:3.4.1}

        Following the detrending process, the key physical parameters of bolometric energy and duration were calculated for each WLF using the resulting detrended optical continuum light curves. As shown in Figure \ref{fig4}(c), prior to calculating the bolometric energy and duration, the detrended optical continuum light curves were interpolated to a uniform cadence of 1 second per data point. This interpolation mitigates potential significant errors in measuring the durations of flares that can arise from the native 45-second cadence of the HMI observations. The bolometric energy of each solar WLF was calculated using the method described in \citet{2013ApJS..209....5S}. The bolometric energy (E) is given by:
        \begin{equation}
        \ E = \sigma_{\mathrm{SB}}T_{\mathrm{flare}}^4\int A_{\mathrm{flare}}(t)dt, \label{eq:4}
        \end{equation}
        \begin{equation}
        \ A_{\mathrm{flare}}(t) = \frac{L_{\mathrm{flare}}}{L_{\mathrm{Sun}}}\pi R^2\frac{\int R_\lambda B_\lambda(5800\mathrm{K})d\lambda}{\int R_\lambda B_\lambda(T_{\mathrm{flare}})d\lambda}, \label{eq:5}
        \end{equation}
        where $\sigma_{\mathrm{SB}}$ represents the Stefan-Boltzmann constant, $T_{\mathrm{flare}} = 10000K$ means that we assume that the WL continuum emission during the flare approximates a blackbody with $T_{\mathrm{flare}} = 10,000K$ \citep{2015SoPh..290.3663K}, $L_{\mathrm{flare}}$/$L_{\mathrm{Sun}}$ stands for the flare luminosity to the overall solar luminosity, $R$ is the solar radius, $B_\lambda(T)$ represents the $Planck$ function at a given wavelength $\lambda$, $R_\lambda$ is a response function of $SDO$/HMI.

        \subsubsection{Refined Bolometric Energy Calculation using Variable-Temperature Blackbody Model}
        \label{subsec:3.4.2}

        Although the method described in Section \ref{subsec:3.4.1} facilitates comparison with earlier studies, assuming a constant temperature of 10,000 K represents an oversimplification, as the temperature evolves significantly during the flare process and varies across the flare kernel \citep{2013ApJ...776..123W}. To address this limitation, we implemented a refined method to calculate the bolometric energy using a time-dependent and pixel-dependent temperature approach. Firstly, we determined the effective background temperature ($T_{bg}$) for each pixel based on its specific location. Pixels within the solar WLFs were categorized into three regions with assigned temperatures: umbra ($T_{bg} \approx 4500$ K), penumbra ($T_{bg} \approx 5500$ K), and quiet region ($T_{bg} \approx 5780$ K). Subsequently, we derived the flare temperature ($T_{flare}(x, y, t)$) for every pixel at each time step ($t$). Considering the narrow bandwidth of the SDO/HMI continuum filter ($\lambda_0 \approx 6173$ Å), we applied a narrowband approximation as expressed in Equation \ref{eq:6}:
        \begin{equation}
        \ I(x,y,t) \propto \int_0^{+\infty} R_{\lambda}B_{\lambda}(T(x,y,t)) d{\lambda} \approx B_{\lambda=6173 \mathring{A}}(T(x,y,t)) \int_{a}^{b} R_{\lambda} d{\lambda}, \label{eq:6}
        \end{equation}
        where the observed intensity is proportional to the product of the Planck function ($B_{\lambda=6173 \mathring{A}}(T_{flare}(x,y,t))$) and the integrated instrument response function ($\int R_{\lambda} d{\lambda}$). By taking the ratio of the flare intensity ($I_{flare}$) to the background intensity ($I_{bg}$), the instrument response function term cancels out. This yields Equation \ref{eq:7}, which directly relates the observed intensity ratio to the ratio of the blackbody radiances:
        \begin{equation}
        \ \frac{I_{flare}(x,y,t)}{I_{bg}(x,y,t)} = \frac{B_{\lambda=6173 \mathring{A}}(T_{flare}(x,y,t))}{B_{\lambda=6173 \mathring{A}}(T_{bg}(x,y,t))}, \label{eq:7}
        \end{equation}
        Finally, with the derived temperatures, we calculated the time profile of the bolometric flux ($F_{bol}(t)$) by integrating the Stefan-Boltzmann emission over the solar WLF area ($S$), , as defined in Equation \ref{eq:8}:
        \begin{equation}
        \ F_{bol}(t) = \int_{S} \sigma_{SB} (T_{flare}(x, y, t)^4) dS, \label{eq:8}
        \end{equation}
        To strictly isolate the bolometric energy released by the flare, we applied a detrending process to this bolometric flux time profile, following the procedures described in Section \ref{subsec:3.3}. The bolometric energy was then obtained by integrating the detrended flux curve over the flare duration, as shown in Equation \ref{eq:9}:
        \begin{equation}
        \ E = \int_{t_{start}}^{t_{end}} (F_{bol}(t) - F_{bg}(t)) dt, \label{eq:9}
        \end{equation}

        Figure \ref{fig5}(a) displays the distributions of bolometric energies calculated using both the fixed-temperature blackbody model ($E_1$) and the refined variable-temperature blackbody model ($E_2$). It is worth noting that the single event displaying anomalously high energy corresponds to the X9.3-class flare in 2017. The comparison clearly indicates that the static fixed-temperature ($10,000$ K) blackbody model tends to overestimate the bolometric energies of solar WLFs. To quantify this discrepancy, Figure \ref{fig5}(b) presents the distribution of the ratio between the two bolometric energies ($E_2 / E_1$). The distribution reveals a mean ratio of 0.23, suggesting that the bolometric energy estimates derived from the traditional fixed-temperature blackbody model are, on average, approximately four times higher than those obtained from the variable-temperature blackbody model. This finding highlights a critical point for existing and future comparative studies between solar and stellar flares: bolometric energy estimations must rigorously account for realistic spatiotemporal temperature variations to avoid significant biases. Neglecting the lower-temperature contributions can lead to greatly overestimated bolometric energy estimates.

        \subsubsection{Duration Calculation}
        \label{subsec:3.4.3}

        Regardless of the method used for the bolometric energy calculation (Section \ref{subsec:3.4.1} or \ref{subsec:3.4.2}), we uniformly calculated the durations of all solar WLFs using the e-folding method \citep{2022ApJ...926L...5N}. Specifically, the rise duration ($t_{rise}$) is defined as the time interval required for the optical continuum light curve or bolometric flux time profile to increase from $1/e$ of its peak intensity to the peak value during the rise phase. Conversely, the decay duration ($t_{decay}$) is defined as the time taken for the emission to decrease from the peak to $1/e$ of the peak value during the decay phase. The total duration ($t_{total}$) is simply the sum of the rise and decay durations: $t_{total} = t_{rise} + t_{decay}$.

\section{Released Data}
\label{sec:released data}

\begin{figure*}[htbp]
    \centering
    \includegraphics [width=0.92\textwidth]{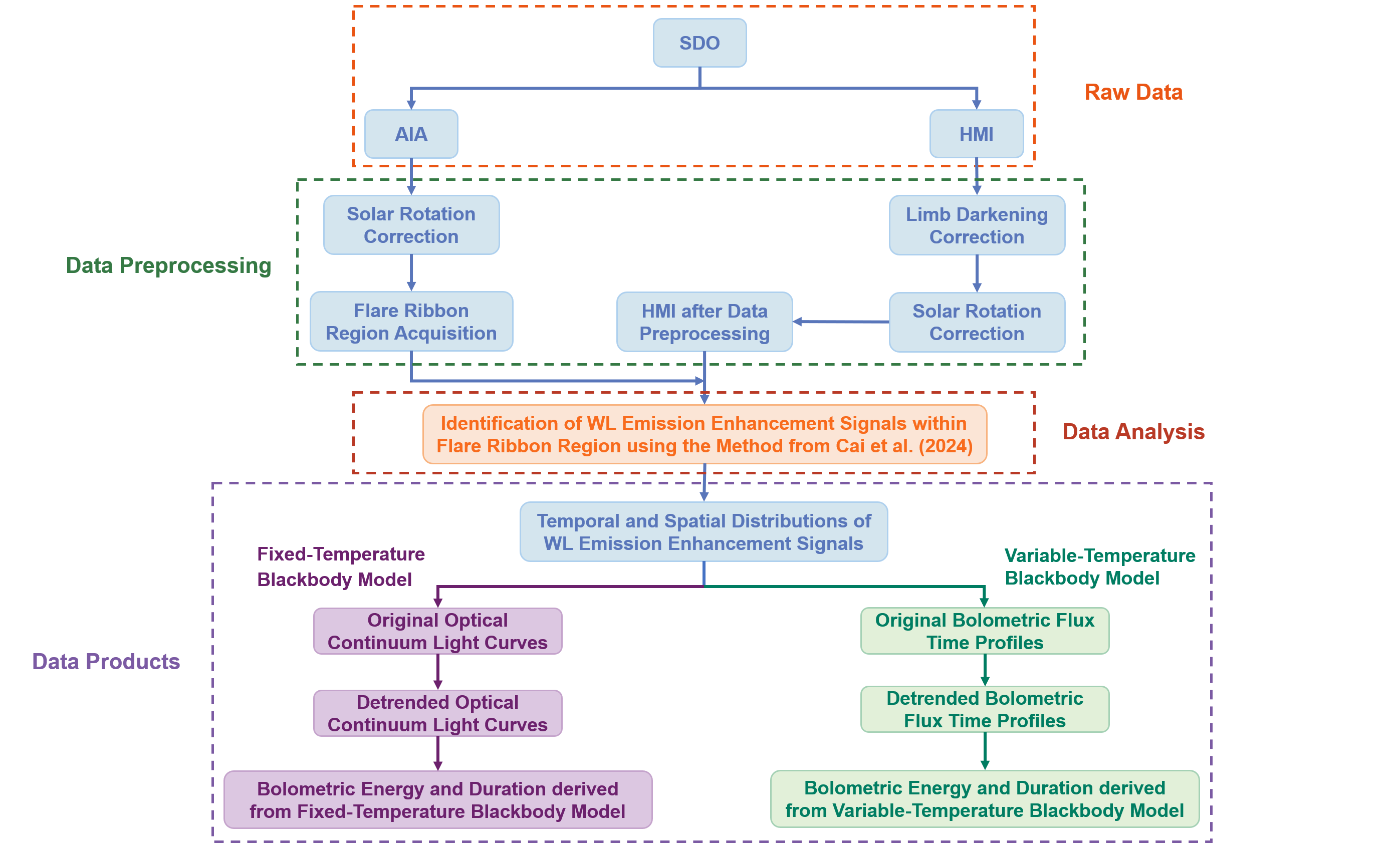}
    \caption{The workflow of constructing the dataset of optical continuum light curves and physical parameters for solar WLFs, including data preprocessing, data analysis and data products.}
    \label{fig6}
    \end{figure*}

    This study releases a comprehensive dataset of 70 solar WLFs (\url{https://doi.org/10.57760/sciencedb.j00167.00040}), which is intended to serve as a foundational resource for the community. Figure \ref{fig6} presents the workflow of constructing the dataset of the optical continuum light curves for solar WLFs, including data preprocessing, data analysis and data products. The dataset is structured into three main components, detailed below.

    \subsection{Locations of White-light Emission Enhancement Signals}
    \label{subsec:4.1}

    This data product provides the precise locations of the identified WL emission enhancement signals for each event. For every solar WLF, we release the following files to fully characterize the spatial data:
    \begin{enumerate}
    \item Preprocessed FITS Images: The calibrated and preprocessed (including limb-darkening and solar rotation correction) HMI 6173 {\AA} continuum intensity images in FITS format.
    \item Pixel Coordinate Tables: A machine-readable table (CSV format) listing the precise pixel coordinates of all identified WL emission enhancement signals within their corresponding preprocessed FITS files during flares.
    \item Quick-Look PNG Images: Supplementary PNG format images that overlay the location of WL emission enhancement signals onto the base HMI 6173 {\AA} continuum intensity images. These provide an immediate visual reference for the spatial distribution and context of the WL emission enhancements.
    \end{enumerate}

    This multi-format approach ensures that users can perform precise quantitative analysis with the FITS data while also benefiting from rapid visual inspection using the PNG images.

    \subsection{Optical Continuum Light Curves of Solar White-light Flares}
    \label{subsec:4.2}

    The second part of the dataset provides the original and detrended optical continuum light curves for each solar WLF. For each event, we release the following files:
    \begin{enumerate}
    \item CSV Tables of optical continuum Light Curves: Machine-readable CSV files containing the original and detrended optical continuum light curves are provided. The timestamps in these files correspond exactly to the observation times of the preprocessed FITS images described in Section \ref{subsec:4.1}.
    \item Quick-Look PNG Images: Supplementary PNG format plots that visually present the optical continuum light curves. These typically display both original and detrended optical continuum light curves for easy verification and presentation.
    \end{enumerate}

    This format provides researchers with both the raw numerical data for rigorous analysis and immediate visualizations for a quick understanding of the flare's temporal characteristics.

    \subsection{Bolometric Energy and Duration of Solar White-light Flares}
    \label{subsec:4.3}

    A summary CSV file is provided, listing the key parameters for each solar WLF, including the flare class, location, and GOES timings (start, peak, end). Crucially, the file provides two distinct sets of calculated bolometric energies and durations, derived from both the fixed-temperature blackbody assumption (Section \ref{subsec:3.4.1}) and the refined variable-temperature blackbody model (Section \ref{subsec:3.4.2}).

\section{Future}
\label{sec:future}
The dataset of 70 solar WLFs presented in this study provides a critical new resource for addressing several questions in solar and stellar flare physics. Two immediate and scientifically promising applications are outlined below.

Due to the scarcity of optical continuum light curves for solar WLFs, statistical studies on the morphology of optical continuum light curve profiles and distributions of bolometric energies and durations for solar WLFs have been scarce. The dataset presented in this work provides a critical resource for addressing this gap, enabling robust, quantitative analyses of these fundamental parameters. Moreover, the energy-duration relation is a key diagnostic metric for flare physics, as it reflects the underlying physical mechanisms of energy release and transport. Our dataset provides a critical supplement to this body of study. Ultimately, the detrended optical continuum light curves make it possible to construct a representative solar WLF template. Such a template describes the characteristic temporal profile of the emission. For stellar flares, a two-component exponential decay template derived from Kepler data \citep{2010Sci...327..977B} has provided key constraints for flare cooling models \citep{2014ApJ...797..122D}. However, the lack of a comparable template for solar WLFs has hindered a direct, like-for-like comparison in the same observational band \citep{2021MNRAS.502.3922K}. Using our dataset, we can now construct the first such template from solar WLFs. Comparing this solar WLF template with established templates from other stars will offer unprecedented insights into the similarities and differences in flare cooling processes between solar and stellar flares.

Beyond these specific analyses, the methodology and pipeline developed here establish a framework for systematic solar WLF studies. Applying this pipeline to the full SDO database will enable the construction of a much larger solar WLF database. This larger sample will be indispensable for studying the solar WLF properties, the mechanisms of  solar WLFs and for performing even more statistically robust comparative studies with stellar WLFs.

\section*{Acknowledgements}

The authors appreciate the anonymous referee for the constructive comments and valuable suggestions. We also appreciate Dr. Xianghan Cui, Dr. Henggeng Han, Changwen Zeng and Houle Huang for their helpful suggestions. The data used here are courtesy of the \emph{SDO} and \emph{GOES} science teams. The authors are supported by the Strategic Priority Research Program of CAS (XDB0560000), the National Key R\&D Program of China (2022YFF0503800), the National Natural Science Foundation of China (12273060, 12588202, 12222306, and 12533010), the Youth Innovation Promotion Association CAS (2023063), China's Space Origins Exploration Program (GJ11020405), and the Specialized Research Fund for State Key Laboratory of Solar Activity and Space Weather. JFL also acknowledges support from the New Cornerstone Science Foundation through the New Cornerstone Investigator Program and the XPLORER PRIZE.

\bibliographystyle{raa}
\bibliography{ref}

@ARTICLE{2024ApJ...975...69C,
       author = {{Cai}, Yingjie and {Hou}, Yijun and {Li}, Ting and {Liu}, Jifeng},
        title = "{Statistics of Solar White-light Flares. I. Optimization and Application of Identification Methods}",
      journal = {\apj},
         year = 2024,
        month = nov,
       volume = {975},
       number = {1},
          eid = {69},
        pages = {69},
          doi = {10.3847/1538-4357/ad793b},
archivePrefix = {arXiv},
       eprint = {2408.05381},
 primaryClass = {astro-ph.SR},
       adsurl = {https://ui.adsabs.harvard.edu/abs/2024ApJ...975...69C}
}

@ARTICLE{2013ApJS..209....5S,
       author = {{Shibayama}, Takuya and {Maehara}, Hiroyuki and {Notsu}, Shota and {Notsu}, Yuta and {Nagao}, Takashi and {Honda}, Satoshi and {Ishii}, Takako T. and {Nogami}, Daisaku and {Shibata}, Kazunari},
        title = "{Superflares on Solar-type Stars Observed with Kepler. I. Statistical Properties of Superflares}",
      journal = {\apjs},
         year = 2013,
        month = nov,
       volume = {209},
       number = {1},
          eid = {5},
        pages = {5},
          doi = {10.1088/0067-0049/209/1/5},
archivePrefix = {arXiv},
       eprint = {1308.1480},
 primaryClass = {astro-ph.SR},
       adsurl = {https://ui.adsabs.harvard.edu/abs/2013ApJS..209....5S}
}

@ARTICLE{2017ApJ...851...91N,
       author = {{Namekata}, Kosuke and {Sakaue}, Takahito and {Watanabe}, Kyoko and {Asai}, Ayumi and {Maehara}, Hiroyuki and {Notsu}, Yuta and {Notsu}, Shota and {Honda}, Satoshi and {Ishii}, Takako T. and {Ikuta}, Kai and {Nogami}, Daisaku and {Shibata}, Kazunari},
        title = "{Statistical Studies of Solar White-light Flares and Comparisons with Superflares on Solar-type Stars}",
      journal = {\apj},
         year = 2017,
        month = dec,
       volume = {851},
       number = {2},
          eid = {91},
        pages = {91},
          doi = {10.3847/1538-4357/aa9b34},
archivePrefix = {arXiv},
       eprint = {1710.11325},
 primaryClass = {astro-ph.SR},
       adsurl = {https://ui.adsabs.harvard.edu/abs/2017ApJ...851...91N}
}

@ARTICLE{1989SoPh..121..261N,
       author = {{Neidig}, Donald F.},
        title = "{The Importance of Solar White-Light Flares}",
      journal = {\solphys},
         year = 1989,
        month = mar,
       volume = {121},
       number = {1-2},
        pages = {261-269},
          doi = {10.1007/BF00161699},
       adsurl = {https://ui.adsabs.harvard.edu/abs/1989SoPh..121..261N}
}

@ARTICLE{1993SoPh..144..169N,
       author = {{Neidig}, Donald F. and {Wiborg}, Philip H. and {Gilliam}, Lou B.},
        title = "{Physical properties of white-light flares derived from their center-to-limb distribution}",
      journal = {\solphys},
         year = 1993,
        month = mar,
       volume = {144},
       number = {1},
        pages = {169-194},
          doi = {10.1007/BF00667990},
       adsurl = {https://ui.adsabs.harvard.edu/abs/1993SoPh..144..169N}
}

@ARTICLE{1995SoPh..157..271F,
       author = {{Fang}, C. and {Henoux}, J.~C. and {Ju}, Hu and {Yin-Zhang}, Xue and {Xiu-Fa}, Gao and {Qi-Jun}, Fu},
        title = "{Semi-Empirical Models of the White-Light Flare on October 24, 1991}",
      journal = {\solphys},
         year = 1995,
        month = mar,
       volume = {157},
       number = {1-2},
        pages = {271-283},
          doi = {10.1007/BF00680621},
       adsurl = {https://ui.adsabs.harvard.edu/abs/1995SoPh..157..271F}
}

@ARTICLE{2003A&A...403.1151D,
       author = {{Ding}, M.~D. and {Liu}, Y. and {Yeh}, C.-T. and {Li}, J.~P.},
        title = "{Interpretation of the infrared continuum in a solar white-light flare}",
      journal = {\aap},
         year = 2003,
        month = jun,
       volume = {403},
        pages = {1151-1156},
          doi = {10.1051/0004-6361:20030428},
       adsurl = {https://ui.adsabs.harvard.edu/abs/2003A&A...403.1151D}
}

@ARTICLE{2017NatCo...8.2202H,
       author = {{Hao}, Q. and {Yang}, K. and {Cheng}, X. and {Guo}, Y. and {Fang}, C. and {Ding}, M.~D. and {Chen}, P.~F. and {Li}, Z.},
        title = "{A circular white-light flare with impulsive and gradual white-light kernels}",
      journal = {Nature Communications},
         year = 2017,
        month = dec,
       volume = {8},
          eid = {2202},
        pages = {2202},
          doi = {10.1038/s41467-017-02343-0},
archivePrefix = {arXiv},
       eprint = {1712.07279},
 primaryClass = {astro-ph.SR},
       adsurl = {https://ui.adsabs.harvard.edu/abs/2017NatCo...8.2202H}
}

@ARTICLE{2018A&A...613A..69S,
       author = {{Song}, Y.~L. and {Tian}, H. and {Zhang}, M. and {Ding}, M.~D.},
        title = "{Observations of white-light flares in NOAA active region 11515: high occurrence rate and relationship with magnetic transients}",
      journal = {\aap},
         year = 2018,
        month = jun,
       volume = {613},
          eid = {A69},
        pages = {A69},
          doi = {10.1051/0004-6361/201731817},
archivePrefix = {arXiv},
       eprint = {1801.04371},
 primaryClass = {astro-ph.SR},
       adsurl = {https://ui.adsabs.harvard.edu/abs/2018A&A...613A..69S}
}

@ARTICLE{2018ApJ...867..159S,
       author = {{Song}, Yongliang and {Tian}, Hui},
        title = "{Investigation of White-light Emission in Circular-ribbon Flares}",
      journal = {\apj},
         year = 2018,
        month = nov,
       volume = {867},
       number = {2},
          eid = {159},
        pages = {159},
          doi = {10.3847/1538-4357/aae5d1},
archivePrefix = {arXiv},
       eprint = {1810.02958},
 primaryClass = {astro-ph.SR},
       adsurl = {https://ui.adsabs.harvard.edu/abs/2018ApJ...867..159S}
}

@ARTICLE{2020ApJ...893L..13S,
       author = {{Song}, Yongliang and {Tian}, Hui and {Zhu}, Xiaoshuai and {Chen}, Yajie and {Zhang}, Mei and {Zhang}, Jingwen},
        title = "{A White-light Flare Powered by Magnetic Reconnection in the Lower Solar Atmosphere}",
      journal = {\apjl},
         year = 2020,
        month = apr,
       volume = {893},
       number = {1},
          eid = {L13},
        pages = {L13},
          doi = {10.3847/2041-8213/ab83fa},
archivePrefix = {arXiv},
       eprint = {2003.11747},
 primaryClass = {astro-ph.SR},
       adsurl = {https://ui.adsabs.harvard.edu/abs/2020ApJ...893L..13S}
}

@ARTICLE{2023ApJ...952L...6S,
       author = {{Song}, De-Chao and {Tian}, Jun and {Li}, Y. and {Ding}, M.~D. and {Su}, Yang and {Yu}, Sijie and {Hong}, Jie and {Qiu}, Ye and {Rao}, Shihao and {Liu}, Xiaofeng and {Li}, Qiao and {Chen}, Xingyao and {Li}, Chuan and {Fang}, Cheng},
        title = "{Spectral Observations and Modeling of a Solar White-light Flare Observed by CHASE}",
      journal = {\apjl},
         year = 2023,
        month = jul,
       volume = {952},
       number = {1},
          eid = {L6},
        pages = {L6},
          doi = {10.3847/2041-8213/ace18c},
archivePrefix = {arXiv},
       eprint = {2307.12641},
 primaryClass = {astro-ph.SR},
       adsurl = {https://ui.adsabs.harvard.edu/abs/2023ApJ...952L...6S}
}

@ARTICLE{2023ApJ...954....7L,
       author = {{Li}, Dong and {Li}, Chuan and {Qiu}, Ye and {Rao}, Shihao and {Warmuth}, Alexander and {Schuller}, Frederic and {Zhao}, Haisheng and {Shi}, Fanpeng and {Xu}, Jun and {Ning}, Zongjun},
        title = "{Observational Signatures of Electron-driven Chromospheric Evaporation in a White-light Flare}",
      journal = {\apj},
         year = 2023,
        month = sep,
       volume = {954},
       number = {1},
          eid = {7},
        pages = {7},
          doi = {10.3847/1538-4357/ace256},
archivePrefix = {arXiv},
       eprint = {2306.15888},
 primaryClass = {astro-ph.SR},
       adsurl = {https://ui.adsabs.harvard.edu/abs/2023ApJ...954....7L}
}

@ARTICLE{2025ApJ...979L..43Y,
       author = {{Yang}, Xu and {Wang}, Meiqi and {Cao}, Andrew and {Ji}, Kaifan and {Yurchyshyn}, Vasyl and {Qiu}, Jiong and {Yu}, Sijie and {Shen}, Jinhua and {Cao}, Wenda},
        title = "{High-resolution Observations of an X-1.0 White-light Flare with Moving Flare Ribbons}",
      journal = {\apjl},
         year = 2025,
        month = feb,
       volume = {979},
       number = {2},
          eid = {L43},
        pages = {L43},
          doi = {10.3847/2041-8213/ada9e4},
       adsurl = {https://ui.adsabs.harvard.edu/abs/2025ApJ...979L..43Y}
}

@ARTICLE{2025ApJ...986L..15X,
       author = {{Xu}, Zhe and {Yan}, Xiaoli and {Li}, Zhentong and {Yang}, Liheng and {Xue}, Zhike and {Wang}, Jincheng and {Zhou}, Yian},
        title = "{High-resolution Observations of a C9.3 White-light Flare and Its Impact on the Solar Photosphere}",
      journal = {\apjl},
         year = 2025,
        month = jun,
       volume = {986},
       number = {1},
          eid = {L15},
        pages = {L15},
          doi = {10.3847/2041-8213/adddb2},
archivePrefix = {arXiv},
       eprint = {2506.08411},
 primaryClass = {astro-ph.SR},
       adsurl = {https://ui.adsabs.harvard.edu/abs/2025ApJ...986L..15X}
}

@ARTICLE{2024ApJ...963L...3L,
       author = {{Li}, Ying and {Jing}, Zhichen and {Song}, De-Chao and {Li}, Qiao and {Tian}, Jun and {Liu}, Xiaofeng and {Wang}, Ya and {Ding}, M.~D. and {Battaglia}, Andrea Francesco and {Feng}, Li and {Li}, Hui and {Gan}, Weiqun},
        title = "{The White-light Emissions in Two X-class Flares Observed by ASO-S and CHASE}",
      journal = {\apjl},
         year = 2024,
        month = mar,
       volume = {963},
       number = {1},
          eid = {L3},
        pages = {L3},
          doi = {10.3847/2041-8213/ad27ca},
archivePrefix = {arXiv},
       eprint = {2402.07374},
 primaryClass = {astro-ph.SR},
       adsurl = {https://ui.adsabs.harvard.edu/abs/2024ApJ...963L...3L}
}

@ARTICLE{2025ApJ...992...72J,
       author = {{Jing}, Zhichen and {Li}, Ying and {Li}, Jingwei and {Li}, Qiao},
        title = "{The M- and X-class White-light Flares in Super Active Region NOAA 13664/13697 Observed by ASO-S/LST/WST}",
      journal = {\apj},
         year = 2025,
        month = oct,
       volume = {992},
       number = {1},
          eid = {72},
        pages = {72},
          doi = {10.3847/1538-4357/ae0708},
archivePrefix = {arXiv},
       eprint = {2509.11029},
 primaryClass = {astro-ph.SR},
       adsurl = {https://ui.adsabs.harvard.edu/abs/2025ApJ...992...72J}
}

@ARTICLE{1970SoPh...13..471S,
       author = {{{\v{S}}vestka}, Z.},
        title = "{The Phase of Particle Acceleration in the Flare Development}",
      journal = {\solphys},
         year = 1970,
        month = aug,
       volume = {13},
       number = {2},
        pages = {471-489},
          doi = {10.1007/BF00153567},
       adsurl = {https://ui.adsabs.harvard.edu/abs/1970SoPh...13..471S}
}

@ARTICLE{2012Natur.485..478M,
       author = {{Maehara}, Hiroyuki and {Shibayama}, Takuya and {Notsu}, Shota and {Notsu}, Yuta and {Nagao}, Takashi and {Kusaba}, Satoshi and {Honda}, Satoshi and {Nogami}, Daisaku and {Shibata}, Kazunari},
        title = "{Superflares on solar-type stars}",
      journal = {\nat},
         year = 2012,
        month = may,
       volume = {485},
       number = {7399},
        pages = {478-481},
          doi = {10.1038/nature11063},
       adsurl = {https://ui.adsabs.harvard.edu/abs/2012Natur.485..478M}
}

@ARTICLE{2010AN....331..636C,
       author = {{Cao}, W. and {Gorceix}, N. and {Coulter}, R. and {Ahn}, K. and {Rimmele}, T.~R. and {Goode}, P.~R.},
        title = "{Scientific instrumentation for the 1.6 m New Solar Telescope in Big Bear}",
      journal = {Astronomische Nachrichten},
         year = 2010,
        month = jun,
       volume = {331},
       number = {6},
        pages = {636},
          doi = {10.1002/asna.201011390},
       adsurl = {https://ui.adsabs.harvard.edu/abs/2010AN....331..636C}
}

@ARTICLE{2014RAA....14..705L,
       author = {{Liu}, Zhong and {Xu}, Jun and {Gu}, Bo-Zhong and {Wang}, Sen and {You}, Jian-Qi and {Shen}, Long-Xiang and {Lu}, Ru-Wei and {Jin}, Zhen-Yu and {Chen}, Lin-Fei and {Lou}, Ke and {Li}, Zhi and {Liu}, Guang-Qian and {Xu}, Zhi and {Rao}, Chang-Hui and {Hu}, Qi-Qian and {Li}, Ru-Feng and {Fu}, Hao-Wen and {Wang}, Feng and {Bao}, Men-Xian and {Wu}, Ming-Chan and {Zhang}, Bo-Rong},
        title = "{New vacuum solar telescope and observations with high resolution}",
      journal = {Research in Astronomy and Astrophysics},
         year = 2014,
        month = jun,
       volume = {14},
       number = {6},
          eid = {705-718},
        pages = {705-718},
          doi = {10.1088/1674-4527/14/6/009},
archivePrefix = {arXiv},
       eprint = {1403.6896},
 primaryClass = {astro-ph.IM},
       adsurl = {https://ui.adsabs.harvard.edu/abs/2014RAA....14..705L}
}

@ARTICLE{2012SoPh..275....3P,
       author = {{Pesnell}, W. Dean and {Thompson}, B.~J. and {Chamberlin}, P.~C.},
        title = "{The Solar Dynamics Observatory (SDO)}",
      journal = {\solphys},
         year = 2012,
        month = jan,
       volume = {275},
       number = {1-2},
        pages = {3-15},
          doi = {10.1007/s11207-011-9841-3},
       adsurl = {https://ui.adsabs.harvard.edu/abs/2012SoPh..275....3P}
}

@ARTICLE{2022SCPMA..6589602L,
       author = {{Li}, Chuan and {Fang}, Cheng and {Li}, Zhen and {Ding}, MingDe and {Chen}, PengFei and {Qiu}, Ye and {You}, Wei and {Yuan}, Yuan and {An}, MinJie and {Tao}, HongJiang and {Li}, XianSheng and {Chen}, Zhe and {Liu}, Qiang and {Mei}, Gui and {Yang}, Liang and {Zhang}, Wei and {Cheng}, WeiQiang and {Chen}, JianXin and {Chen}, ChangYa and {Gu}, Qiang and {Huang}, QingLong and {Liu}, MingXing and {Han}, ChengShan and {Xin}, HongWei and {Chen}, ChangZheng and {Ni}, YiWei and {Wang}, WenBo and {Rao}, ShiHao and {Li}, HaiTang and {Lu}, Xi and {Wang}, Wei and {Lin}, Jun and {Jiang}, YiXian and {Meng}, LingJie and {Zhao}, Jian},
        title = "{The Chinese H{\ensuremath{\alpha}} Solar Explorer (CHASE) mission: An overview}",
      journal = {Science China Physics, Mechanics, and Astronomy},
         year = 2022,
        month = aug,
       volume = {65},
       number = {8},
          eid = {289602},
        pages = {289602},
          doi = {10.1007/s11433-022-1893-3},
archivePrefix = {arXiv},
       eprint = {2205.05962},
 primaryClass = {astro-ph.SR},
       adsurl = {https://ui.adsabs.harvard.edu/abs/2022SCPMA..6589602L}
}

@ARTICLE{2023SoPh..298...68G,
       author = {{Gan}, Weiqun and {Zhu}, Cheng and {Deng}, Yuanyong and {Zhang}, Zhe and {Chen}, Bo and {Huang}, Yu and {Deng}, Lei and {Wu}, Haiyan and {Zhang}, Haiying and {Li}, Hui and {Su}, Yang and {Su}, Jiangtao and {Feng}, Li and {Wu}, Jian and {Cui}, Jijun and {Wang}, Chi and {Chang}, Jin and {Yin}, Zengshan and {Xiong}, Weiming and {Chen}, Bin and {Yang}, Jianfeng and {Li}, Fu and {Lin}, Jiaben and {Hou}, Junfeng and {Bai}, Xianyong and {Chen}, Dengyi and {Zhang}, Yan and {Hu}, Yiming and {Liang}, Yaoming and {Wang}, Jianping and {Song}, Kefei and {Guo}, Quanfeng and {He}, Lingping and {Zhang}, Guang and {Wang}, Peng and {Bao}, Haicao and {Cao}, Caixia and {Bai}, Yanping and {Chen}, Binglong and {He}, Tao and {Li}, Xinyu and {Zhang}, Ye and {Liao}, Xing and {Jiang}, Hu and {Li}, Youping and {Su}, Yingna and {Lei}, Shijun and {Chen}, Wei and {Li}, Ying and {Zhao}, Jie and {Li}, Jingwei and {Ge}, Yunyi and {Zou}, Ziming and {Hu}, Tai and {Su}, Miao and {Ji}, Haidong and {Gu}, Mei and {Zheng}, Yonghuang and {Xu}, Dezhen and {Wang}, Xing},
        title = "{The Advanced Space-Based Solar Observatory (ASO-S)}",
      journal = {\solphys},
         year = 2023,
        month = may,
       volume = {298},
       number = {5},
          eid = {68},
        pages = {68},
          doi = {10.1007/s11207-023-02166-x},
       adsurl = {https://ui.adsabs.harvard.edu/abs/2023SoPh..298...68G}
}

@ARTICLE{2012SoPh..275..207S,
       author = {{Scherrer}, P.~H. and {Schou}, J. and {Bush}, R.~I. and {Kosovichev}, A.~G. and {Bogart}, R.~S. and {Hoeksema}, J.~T. and {Liu}, Y. and {Duvall}, T.~L. and {Zhao}, J. and {Title}, A.~M. and {Schrijver}, C.~J. and {Tarbell}, T.~D. and {Tomczyk}, S.},
        title = "{The Helioseismic and Magnetic Imager (HMI) Investigation for the Solar Dynamics Observatory (SDO)}",
      journal = {\solphys},
         year = 2012,
        month = jan,
       volume = {275},
       number = {1-2},
        pages = {207-227},
          doi = {10.1007/s11207-011-9834-2},
       adsurl = {https://ui.adsabs.harvard.edu/abs/2012SoPh..275..207S}
}

@ARTICLE{2012SoPh..275...17L,
       author = {{Lemen}, James R. and {Title}, Alan M. and {Akin}, David J. and {Boerner}, Paul F. and {Chou}, Catherine and {Drake}, Jerry F. and {Duncan}, Dexter W. and {Edwards}, Christopher G. and {Friedlaender}, Frank M. and {Heyman}, Gary F. and {Hurlburt}, Neal E. and {Katz}, Noah L. and {Kushner}, Gary D. and {Levay}, Michael and {Lindgren}, Russell W. and {Mathur}, Dnyanesh P. and {McFeaters}, Edward L. and {Mitchell}, Sarah and {Rehse}, Roger A. and {Schrijver}, Carolus J. and {Springer}, Larry A. and {Stern}, Robert A. and {Tarbell}, Theodore D. and {Wuelser}, Jean-Pierre and {Wolfson}, C. Jacob and {Yanari}, Carl and {Bookbinder}, Jay A. and {Cheimets}, Peter N. and {Caldwell}, David and {Deluca}, Edward E. and {Gates}, Richard and {Golub}, Leon and {Park}, Sang and {Podgorski}, William A. and {Bush}, Rock I. and {Scherrer}, Philip H. and {Gummin}, Mark A. and {Smith}, Peter and {Auker}, Gary and {Jerram}, Paul and {Pool}, Peter and {Soufli}, Regina and {Windt}, David L. and {Beardsley}, Sarah and {Clapp}, Matthew and {Lang}, James and {Waltham}, Nicholas},
        title = "{The Atmospheric Imaging Assembly (AIA) on the Solar Dynamics Observatory (SDO)}",
      journal = {\solphys},
         year = 2012,
        month = jan,
       volume = {275},
       number = {1-2},
        pages = {17-40},
          doi = {10.1007/s11207-011-9776-8},
       adsurl = {https://ui.adsabs.harvard.edu/abs/2012SoPh..275...17L}
}

@ARTICLE{2020ApJ...904...96C,
       author = {{Castellanos Dur{\'a}n}, J. Sebasti{\'a}n and {Kleint}, Lucia},
        title = "{The Statistical Relationship between White-light Emission and Photospheric Magnetic Field Changes in Flares}",
      journal = {\apj},
         year = 2020,
        month = dec,
       volume = {904},
       number = {2},
          eid = {96},
        pages = {96},
          doi = {10.3847/1538-4357/ab9c1e},
archivePrefix = {arXiv},
       eprint = {2007.02954},
 primaryClass = {astro-ph.SR},
       adsurl = {https://ui.adsabs.harvard.edu/abs/2020ApJ...904...96C}
}

@BOOK{1976asqu.book.....A,
       author = {{Allen}, C.~W.},
        title = "{Astrophysical Quantities}",
         year = 1976,
       adsurl = {https://ui.adsabs.harvard.edu/abs/1976asqu.book.....A}
}

@ARTICLE{2014ApJ...797..122D,
       author = {{Davenport}, James R.~A. and {Hawley}, Suzanne L. and {Hebb}, Leslie and {Wisniewski}, John P. and {Kowalski}, Adam F. and {Johnson}, Emily C. and {Malatesta}, Michael and {Peraza}, Jesus and {Keil}, Marcus and {Silverberg}, Steven M. and {Jansen}, Tiffany C. and {Scheffler}, Matthew S. and {Berdis}, Jodi R. and {Larsen}, Daniel M. and {Hilton}, Eric J.},
        title = "{Kepler Flares. II. The Temporal Morphology of White-light Flares on GJ 1243}",
      journal = {\apj},
         year = 2014,
        month = dec,
       volume = {797},
       number = {2},
          eid = {122},
        pages = {122},
          doi = {10.1088/0004-637X/797/2/122},
archivePrefix = {arXiv},
       eprint = {1411.3723},
 primaryClass = {astro-ph.SR},
       adsurl = {https://ui.adsabs.harvard.edu/abs/2014ApJ...797..122D}
}

@ARTICLE{2015SoPh..290.3663K,
       author = {{Katsova}, M.~M. and {Livshits}, M.~A.},
        title = "{The Origin of Superflares on G-Type Dwarf Stars of Various Ages}",
      journal = {\solphys},
         year = 2015,
        month = dec,
       volume = {290},
       number = {12},
        pages = {3663-3682},
          doi = {10.1007/s11207-015-0752-6},
archivePrefix = {arXiv},
       eprint = {1508.00254},
 primaryClass = {astro-ph.SR},
       adsurl = {https://ui.adsabs.harvard.edu/abs/2015SoPh..290.3663K}
}

@ARTICLE{2022ApJ...926L...5N,
       author = {{Namekata}, Kosuke and {Maehara}, Hiroyuki and {Honda}, Satoshi and {Notsu}, Yuta and {Okamoto}, Soshi and {Takahashi}, Jun and {Takayama}, Masaki and {Ohshima}, Tomohito and {Saito}, Tomoki and {Katoh}, Noriyuki and {Tozuka}, Miyako and {Murata}, Katsuhiro L. and {Ogawa}, Futa and {Niwano}, Masafumi and {Adachi}, Ryo and {Oeda}, Motoki and {Shiraishi}, Kazuki and {Isogai}, Keisuke and {Nogami}, Daisaku and {Shibata}, Kazunari},
        title = "{Discovery of a Long-duration Superflare on a Young Solar-type Star EK Draconis with Nearly Similar Time Evolution for H{\ensuremath{\alpha}} and White-light Emissions}",
      journal = {\apjl},
         year = 2022,
        month = feb,
       volume = {926},
       number = {1},
          eid = {L5},
        pages = {L5},
          doi = {10.3847/2041-8213/ac4df0},
archivePrefix = {arXiv},
       eprint = {2201.09416},
 primaryClass = {astro-ph.SR},
       adsurl = {https://ui.adsabs.harvard.edu/abs/2022ApJ...926L...5N}
}

@ARTICLE{2010Sci...327..977B,
       author = {{Borucki}, William J. and {Koch}, David and {Basri}, Gibor and {Batalha}, Natalie and {Brown}, Timothy and {Caldwell}, Douglas and {Caldwell}, John and {Christensen-Dalsgaard}, J{\o}rgen and {Cochran}, William D. and {DeVore}, Edna and {Dunham}, Edward W. and {Dupree}, Andrea K. and {Gautier}, Thomas N. and {Geary}, John C. and {Gilliland}, Ronald and {Gould}, Alan and {Howell}, Steve B. and {Jenkins}, Jon M. and {Kondo}, Yoji and {Latham}, David W. and {Marcy}, Geoffrey W. and {Meibom}, S{\o}ren and {Kjeldsen}, Hans and {Lissauer}, Jack J. and {Monet}, David G. and {Morrison}, David and {Sasselov}, Dimitar and {Tarter}, Jill and {Boss}, Alan and {Brownlee}, Don and {Owen}, Toby and {Buzasi}, Derek and {Charbonneau}, David and {Doyle}, Laurance and {Fortney}, Jonathan and {Ford}, Eric B. and {Holman}, Matthew J. and {Seager}, Sara and {Steffen}, Jason H. and {Welsh}, William F. and {Rowe}, Jason and {Anderson}, Howard and {Buchhave}, Lars and {Ciardi}, David and {Walkowicz}, Lucianne and {Sherry}, William and {Horch}, Elliott and {Isaacson}, Howard and {Everett}, Mark E. and {Fischer}, Debra and {Torres}, Guillermo and {Johnson}, John Asher and {Endl}, Michael and {MacQueen}, Phillip and {Bryson}, Stephen T. and {Dotson}, Jessie and {Haas}, Michael and {Kolodziejczak}, Jeffrey and {Van Cleve}, Jeffrey and {Chandrasekaran}, Hema and {Twicken}, Joseph D. and {Quintana}, Elisa V. and {Clarke}, Bruce D. and {Allen}, Christopher and {Li}, Jie and {Wu}, Haley and {Tenenbaum}, Peter and {Verner}, Ekaterina and {Bruhweiler}, Frederick and {Barnes}, Jason and {Prsa}, Andrej},
        title = "{Kepler Planet-Detection Mission: Introduction and First Results}",
      journal = {Science},
         year = 2010,
        month = feb,
       volume = {327},
       number = {5968},
        pages = {977},
          doi = {10.1126/science.1185402},
       adsurl = {https://ui.adsabs.harvard.edu/abs/2010Sci...327..977B}
}

@ARTICLE{2021MNRAS.502.3922K,
       author = {{Kashapova}, Larisa K. and {Broomhall}, Anne-Marie and {Larionova}, Alena I. and {Kupriyanova}, Elena G. and {Motyk}, Ilya D.},
        title = "{The morphology of average solar flare time profiles from observations of the Sun's lower atmosphere}",
      journal = {\mnras},
         year = 2021,
        month = apr,
       volume = {502},
       number = {3},
        pages = {3922-3931},
          doi = {10.1093/mnras/stab276},
archivePrefix = {arXiv},
       eprint = {2102.02596},
 primaryClass = {astro-ph.SR},
       adsurl = {https://ui.adsabs.harvard.edu/abs/2021MNRAS.502.3922K}
}

@ARTICLE{2024SoPh..299...11J,
       author = {{Jing}, Zhichen and {Li}, Ying and {Feng}, Li and {Li}, Hui and {Huang}, Yu and {Li}, Youping and {Su}, Yang and {Chen}, Wei and {Tian}, Jun and {Song}, Dechao and {Li}, Jingwei and {Xue}, Jianchao and {Zhao}, Jie and {Lu}, Lei and {Ying}, Beili and {Zhang}, Ping and {Su}, Yingna and {Zhang}, Qingmin and {Li}, Dong and {Ge}, Yunyi and {Li}, Shuting and {Li}, Qiao and {Li}, Gen and {Liu}, Xiaofeng and {Shi}, Guanglu and {Shan}, Jiahui and {Tian}, Zhengyuan and {Zhou}, Yue and {Gan}, Weiqun},
        title = "{A Statistical Study of Solar White-Light Flares Observed by the White-Light Solar Telescope of the Lyman-Alpha Solar Telescope on the Advanced Space-Based Solar Observatory (ASO-S/LST/WST) at 360 nm}",
      journal = {\solphys},
         year = 2024,
        month = jan,
       volume = {299},
       number = {2},
          eid = {11},
        pages = {11},
          doi = {10.1007/s11207-024-02251-9},
archivePrefix = {arXiv},
       eprint = {2401.07275},
 primaryClass = {astro-ph.SR},
       adsurl = {https://ui.adsabs.harvard.edu/abs/2024SoPh..299...11J}
}

@ARTICLE{2013ApJ...776..123W,
       author = {{Watanabe}, Kyoko and {Shimizu}, Toshifumi and {Masuda}, Satoshi and {Ichimoto}, Kiyoshi and {Ohno}, Masanori},
        title = "{Emission Height and Temperature Distribution of White-light Emission Observed by Hinode/SOT from the 2012 January 27 X-class Solar Flare}",
      journal = {\apj},
         year = 2013,
        month = oct,
       volume = {776},
       number = {2},
          eid = {123},
        pages = {123},
          doi = {10.1088/0004-637X/776/2/123},
archivePrefix = {arXiv},
       eprint = {1308.5059},
 primaryClass = {astro-ph.SR},
       adsurl = {https://ui.adsabs.harvard.edu/abs/2013ApJ...776..123W}
}
\end{document}